\documentclass[aps,prl,twocolumn,showpacs]{revtex4}%
\usepackage{graphicx}
\usepackage{dcolumn}
\usepackage{bm}
\usepackage{amssymb}
\usepackage{amsmath}
\usepackage{amsfonts}%
\providecommand{\U}[1]{\protect\rule{.1in}{.1in}}
\begin{document}
\title{ N\'{e}el Spin Orbit Torque driven antiferromagnetic resonance in Mn$_{2}$Au
probed by time-domain THz spectroscopy}
\author{N. Bhattacharjee,$^{1}$ A.A. Sapozhnik,$^{1,2}$ S.Yu. Bodnar,$^{1}$ V.Yu.
Grigorev,$^{1,2}$ S.Y. Agustsson,$^{1}$ J. Cao,$^{1}$ D. Dominko,$^{1}$ M.
Obergfell,$^{1}$ O. Gomonay,$^{1}$ J. Sinova,$^{1,3}$ M. Kl\"{a}ui,$^{1}$
H.-J. Elmers,$^{1}$ M. Jourdan,$^{1}$ and J. Demsar$^{1}$}
\affiliation{$^{1}$Institute of Physics, Johannes Gutenberg-University Mainz, 55099 Mainz, Germany}
\affiliation{$^{2}$Graduate School of Excellence, Materials Science in Mainz (MAINZ),
Mainz, Germany}
\affiliation{$^{3}$Institute of Physics ASCR, v.v.i., Cukrovarnicka 10, 162 53 Praha 6
Czech Republic}
\date{\today}


\begin{abstract}
We observe the excitation of collective modes in the THz range driven by the
recently discovered N\'{e}el spin-orbit torques (NSOT) in the metallic
antiferromagnet Mn$_{2}$Au. Temperature dependent THz spectroscopy reveals a
strong absorption mode centered near 1 THz, which upon heating from 4 K to 450
K softens and looses intensity. Comparison with the estimated eigenmode
frequencies implies that the observed mode is an in-plane antiferromagnetic
resonance (AFMR) mode. The AFMR absorption strength exceeds those found in
antiferromagnetic insulators, driven by the magnetic field of the THz
radiation, by three orders of magnitude. Based on this and the agreement with
our theory modelling, we infer that the driving mechanism for the observed
mode is the current induced NSOT. This electric manipulation of the Ne\'{e}l
order parameter at high frequencies makes Mn$_{2}$Au a prime candidate for AFM
ultrafast memory applications.

\end{abstract}
\maketitle

Antiferromagnetic (AFM) materials are attracting significant interest in the
field of spintronics \cite{Zutic,MacDonald,Gomonay,Jungwirth,Baltz}. The
magnetic order consisting of alternating magnetic moments on neighboring atoms
results in zero net magnetization and makes the AFMs insensitive to external
magnetic fields. At the same time, it also leads to the absence of stray
fields in AFMs. Compared to ferromagnets, this property can enable a
significant reduction of the minimum volume necessary to store one bit of
information. On the other hand, insensitivity of AFMs to external magnetic
field makes an efficient manipulation and detection of the magnetic state of
an AFM challenging.

A key advance in overcoming this challenge is the recent proposal
\cite{Zelezny} of an all-electrical switching of the staggered magnetization
of metallic AFMs via an electrically induced N\'{e}el spin-orbit torque
(NSOT), which has been experimentally realised recently
\cite{Wadley,Bodnar,Meinert,Saidl}. The NSOT takes place in AFMs with a lack
of local inversion symmetry. Here, charge currents produce a staggered
N\'{e}el spin-orbit torque (NSOT) \cite{Zelezny}, which presents a novel route
of manipulating metallic AFMs. Two materials, CuMnAs \cite{Wadley,Wad} and
Mn$_{2}$Au \cite{Mohn,Barthem,Jourdan}, that meet the indicated requirements
are known to date. In both materials, switching of the N\'{e}el vector by
pulsed DC currents was demonstrated \cite{Wadley,WadleyNew,Bodnar,Meinert}. In
CuMnAs, N\'{e}el vector switching was also realised recently using intense THz
pulses \cite{Olejnik}. Provided that frequencies of collective modes, i.e.
antiferromagnetic resonances (AFMR),\ in these metallic AFMs indeed lie in the
terahertz (THz) range \cite{Tinkham,Kampfrath,THzAFMR}, ultrafast switching of
the N\'{e}el vector on the picosecond timescale can be achieved.

Among the metallic AFM materials Mn$_{2}$Au is of special interest due to its
high N\'{e}el temperature ($\approx1500$ K) \cite{Mohn,Barthem}, strong
spin-orbit coupling, and high conductivity \cite{Jourdan}. Bringing the
recently demonstrated switching of the N\'{e}el vector by pulsed DC currents
\cite{Bodnar,Meinert} towards the THz regime is a necessary key advance for
the possible realisation of ultrafast switching and a bridge materials to
overcome the THz communication gap.

In this Letter, we report on the temperature dependent THz conductivity data
of c-axis epitaxial Mn$_{2}$Au thin films. In addition to the Drude free
carrier response, the complex THz conductivity data reveal the presence of a
mode, centered near 1 THz at 4 K. The mode displays significant softening and
loss of intensity upon increasing the temperature to 450 K, as expected for an
AFMR. By comparing the experimentally determined mode frequency to the
theoretically estimated values we attribute the 1 THz mode to an in-plane
AFMR. The mode's absorption strength is found to be about three orders of
magnitude larger that in typical insulating antiferromagnets, and cannot be
accounted for by the coupling to the magnetic field component of the THz
pulse. By comparing the magnitudes of the Zeeman torque with expected values
of the N\'{e}el spin-orbit torque in Mn$_{2}$Au, we conclude that the in-plane
AFMR mode\ in Mn$_{2}$Au is driven by the AC currents via the NSOT.

\begin{figure}[h]
\includegraphics[width=8.5cm]{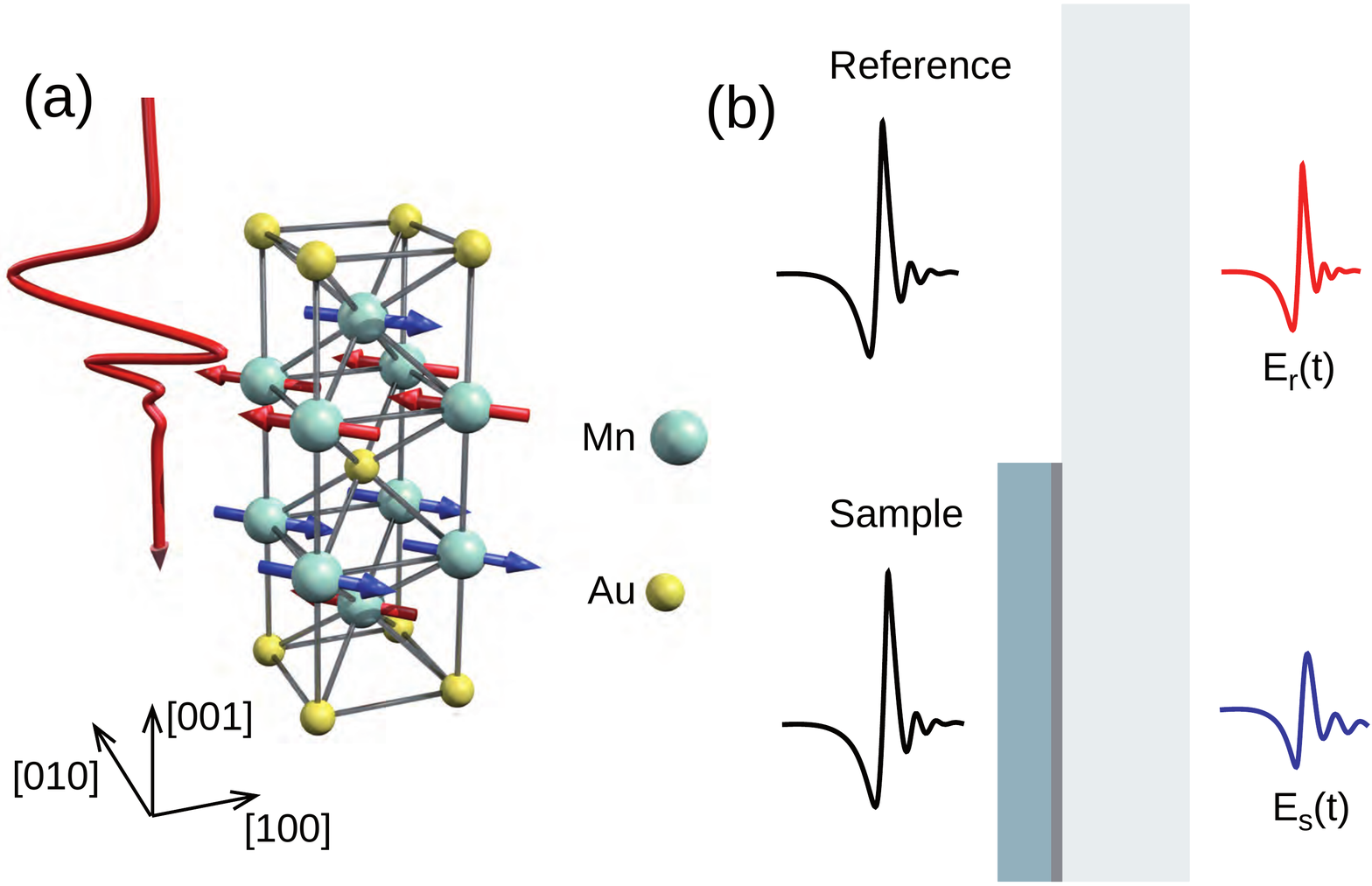}\caption{(a) Lattice and spin structure
of Mn$_{2}$Au. Au and Mn atoms are shown in yellow and violet respectively.
Spin orientation in adjacent layers is given by red and blue arrows, with the
N\'{e}el vector pointing along the [110] direction (note that in the sample
there is roughly an equal amount of domains with the N\'{e}el vector pointing
along $[110]$ and [$1\overline{1}0$]). The THz pulse propagates along the
[001] direction, with polarization parallel or perpendicular to the N\'{e}el
vector. (b) The schematic layout of the TDTS experiment.}%
\end{figure}

We have grown C-axis (001) oriented Mn$_{2}$Au thin films, with the crystal
structure depicted in Fig. 1(a), on a 530 $\mu$m thick r-cut Al$_{2}$O$_{3}$
substrate with the lateral size of 10$\times$10 mm$^{2}$ by radio--frequency
magnetron sputtering \ \cite{Jourdan}. To ensure epitaxial growth, 40 nm thick
Mn$_{2}$Au films are deposited on a 8 nm thick (001) Ta buffer layer. The
films are capped by $\sim2$ nm of Al, forming an insulating aluminum oxide, to
protect Mn$_{2}$Au from oxidation. One half of the metallic film is etched off
to serve as a reference in time-domain THz spectrometer (TDTS) in transmission
configuration (see Fig. 1(b)). The home-built TDTS set-up is built around a
300 kHz amplified Ti:sapphire laser system \cite{beckNbN,BeckPCCO}, utilizing
a large area interdigitated photo-conductive emitter for the generation of THz
pulses \cite{beckOE}. The THz electric field pulses with the peak electric
field strength of $\sim5$ kV$/$cm, are transmitted through the sample,
$E_{\mathrm{s}}(t)$, and the reference, $E_{\mathrm{r}}(t)$, and detected
using the Pockels effect in GaP \cite{beckOE}. THz pulses are polarized along
the\ [110] easy axis of Mn$_{2}$Au (see Fig. 1(a)), which is parallel to the
fast optical axis of the substrate. Thin Ta films, used as a buffer layer for
epitaxial growth of Mn$_{2}$Au, have a nearly identical resistivity to
Mn$_{2}$Au (100 $\mu\Omega$cm at room temperature \cite{Jourdan,Baker}). Thus,
we can treat the thin Ta buffer layer for the following analysis as a part of
the homogeneous Mn$_{2}$Au film. We use the thin film approximation
\cite{THzreview} to extract the complex refractive index, $n\left(
\nu\right)  =\sqrt{\epsilon\left(  \nu\right)  \mu\left(  \nu\right)  }$, from
the optical transfer function \cite{Supp}. Since the refractive index of
Mn$_{2}$Au is largely dominated by the conductive response, and the
permeability of antiferromagnets far below the N\'{e}el temperature is very
close to 1, we analyze the data in terms of the real, $\sigma_{1}(\nu)$, and
the imaginary, $\sigma_{2}(\nu)$, component of the optical conductivity (we
present the extracted $n=\sqrt{\epsilon\mu}$ in the Supplementary Information
\cite{Supp}).

\begin{figure}[h]
\includegraphics[width=8.5cm]{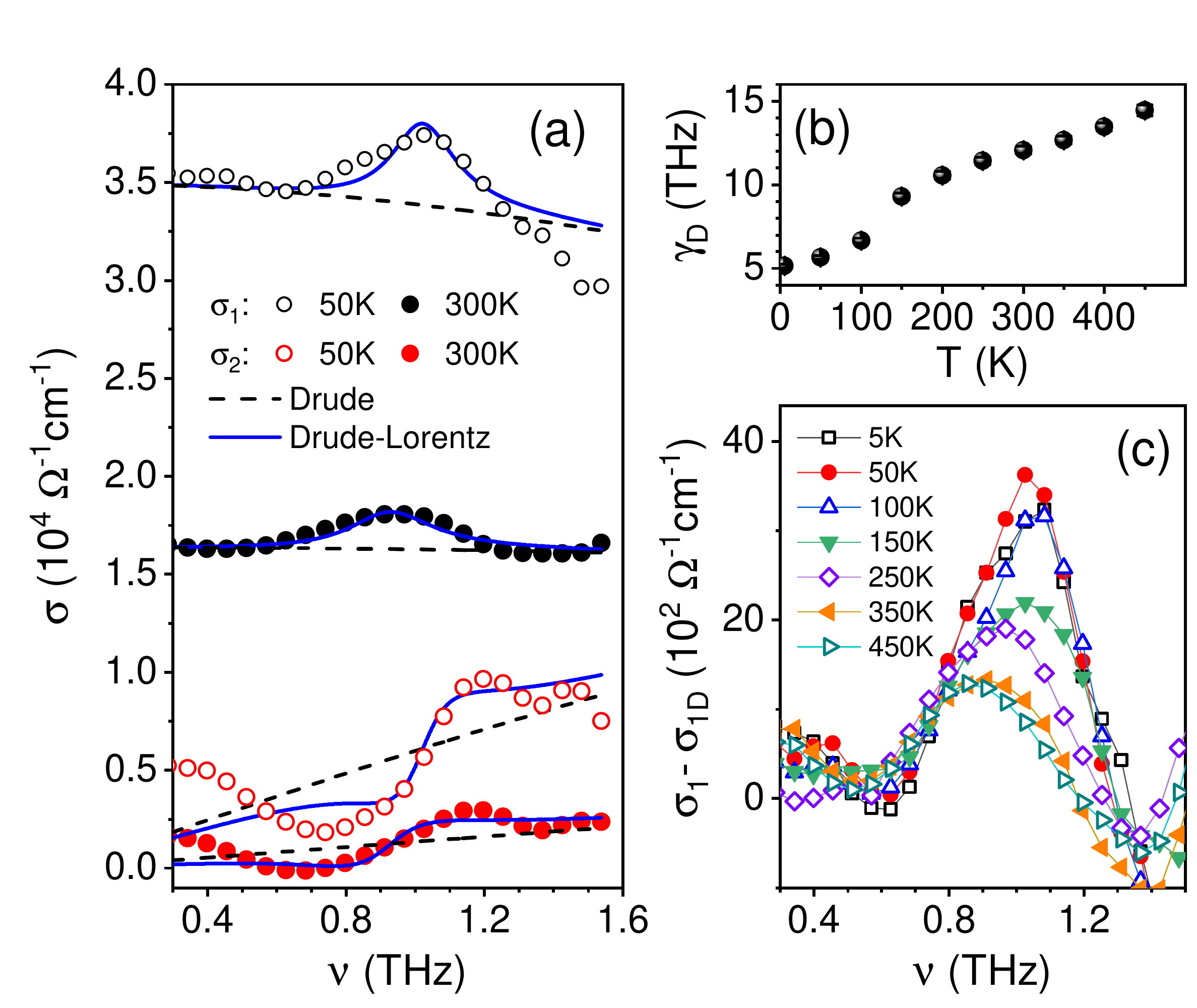}\caption{(color online) (a) The complex
optical conductivity of Mn$_{2}$Au recorded at 50 K and 300 K. Dashed black
lines present the best fits with the Drude model, while the fits with the
Drude-Lorentz model are given by solid blue lines. (b) Temperature evolution
of the real part of the conductivity with the Drude free carrier contribution
subtracted. (c) Temperature dependence of the Drude scattering rate.}%
\end{figure}

The measurements of $\sigma(\nu)$ were performed between 4 K and 450 K. The
real and imaginary part of the THz conductivity, recorded at 50 and 300 K, are
presented in Fig. 2(a). $\sigma(\nu)$ is clearly dominated by the Drude free
carrier response. The low frequency limit of $\sigma_{1}(\nu,$ 300 K$)$
matches the room temperature DC conductivity, recorded on the same sample by
the Van der Pauw method. Dashed black lines in Figure 2(a) present the best
fit to the data obtained by the simple Drude model, $\sigma^{\mathrm{D}}%
(\nu)=2\pi\varepsilon_{0}\nu_{\mathrm{D}}^{2}/(\gamma_{\mathrm{D}}%
-\mathrm{i}\nu)$, with the plasma frequency $\nu_{\mathrm{D}}\approx600$ THz,
and the free carrier scattering rate, $\gamma_{\mathrm{D}}\left(  300\text{
K}\right)  \approx12$ THz. In addition, a distinct spectral feature,
consistent with a resonance centered around $1$ THz, is observed for all
temperatures in both $\sigma_{1}(\nu)$ and $\sigma_{2}(\nu)$. To account for
both, the free carrier response and the $\approx1$ THz mode, we use the
Drude-Lorentz model \cite{Supp} (solid blue lines), where a finite frequency
mode with a linewidth of $\approx0.3$ THz\ is centered near $\approx1$ THz.
There is an upturn in $\sigma_{2}$ at $\nu\lesssim0.5$ THz, which might
suggest a possible second low-frequency mode centered much below $0.2$ THz
\cite{Raman}. Since the use of a 3 mm aperture limits our spectral range to
$\nu>0.3$ THz, we do not include additional modes in the data analysis.

Figure 2(b) presents the temperature dependence of the Drude scattering rate
(plasma frequency was kept constant). It coincides with the temperature
dependence of the DC resistivity \cite{Jourdan}. To track the temperature
dependence of the $\approx1$ THz mode, we subtracted the Drude part,
$\sigma^{\mathrm{D}}(\nu,T)$, from the experimental $\sigma(\nu,T)$. As shown
by $\sigma_{1}(\nu,T)-\sigma_{1}^{\mathrm{D}}(\nu,T)$, presented in Fig. 2(c),
the mode frequency clearly red shifts with increasing temperature. While the
linewidth shows no temperature dependence within the measured temperature
range, the mode's spectral weight, S$_{\text{{\tiny AFMR}}}\varpropto$ $%
{\textstyle\int}
\left(  \sigma_{1}(\nu,T)-\sigma_{1}^{\mathrm{D}}(\nu,T)\right)  d\nu$, is
substantially reduced upon increasing the temperature. 

\begin{figure}[h]
\includegraphics[width=8.5cm]{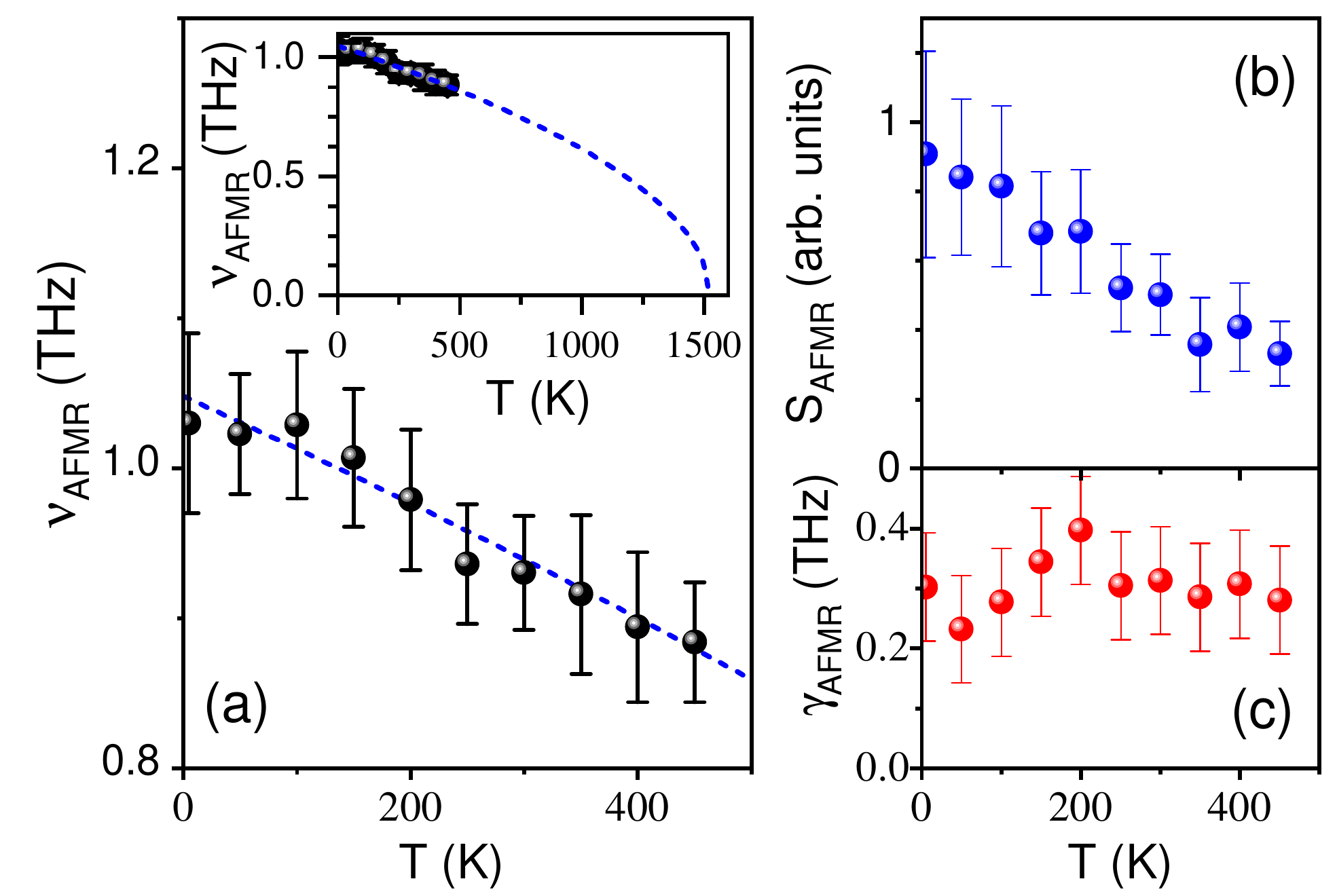}\caption{ (color online) (a)
Temperature dependence of the AFMR frequency shows a significant softening
upon heating from 4 to 450 K. Fitting the T-dependence with the
Ginzburg-Landau mean field expression (dashed blue line) suggests the N\'{e}el
temperature around 1500 K (inset).\ (b) The mode's spectral weight,
$S_{\text{{\protect\tiny AFMR}}}$, displays a three-fold reduction upon
increasing the temperature to 450 K. (c) The extracted linewidth,
$\gamma_{\text{{\protect\tiny AFMR}}}(T)$, shows no T-dependence variation
within the given uncertainty. }%
\end{figure}

Based on the low frequency of the mode and its temperature
dependence we attribute the mode to one of the two $q=0$ antiferromagnetic
eigenmodes of Mn$_{2}$Au. The parameters of this AFMR, obtained by analyzing
$\sigma(\nu,T)$ with the Drude-Lorentz model, are presented in Fig. 3. The
mode frequency, $\nu_{\text{{\tiny AFMR}}}$, clearly shows a shift from 1.03
THz to 0.88 THz as the temperature increases from 5 K to 450 K (Fig. 3(a)).
For Mn$_{2}$Au only the extrapolated value of the N\'{e}el temperature,
$T_{\mathrm{n}}$, is known (1300 K $<T_{\mathrm{n}}<1600$ K) \cite{Barthem}%
,\ since the crystal becomes structurally unstable around 1000 K. Furthermore,
little is known of the temperature dependence of anisotropy fields in Mn$_{2}%
$Au. Thus, we use a simple Ginzburg-Landau model, $\nu_{\text{{\tiny AFMR}}%
}\varpropto\sqrt{T_{\mathrm{n}}-T}$, to evaluate the temperature dependence of
$\nu_{\text{{\tiny AFMR}}}$. As shown in the inset of Fig. 3(a), the resulting
extrapolated value of $T_{\mathrm{n}}$ is 1500 K, which is consistent with
previously reported estimates \cite{Barthem}. We observe a pronounced loss of
the mode's spectral weight, S$_{\text{{\tiny AFMR}}}$, signifying a gradual
decrease of the order parameter (i.e., the mode should disappear at
$T_{\mathrm{n}}$). The mode damping, $\gamma_{\text{{\tiny AFMR}}}$, shows no
measurable temperature dependence, suggesting that the damping is due to the
interaction with free carriers which are mostly unaffected in this temperature range.

\begin{figure}[h]
\includegraphics[width=8.5cm]{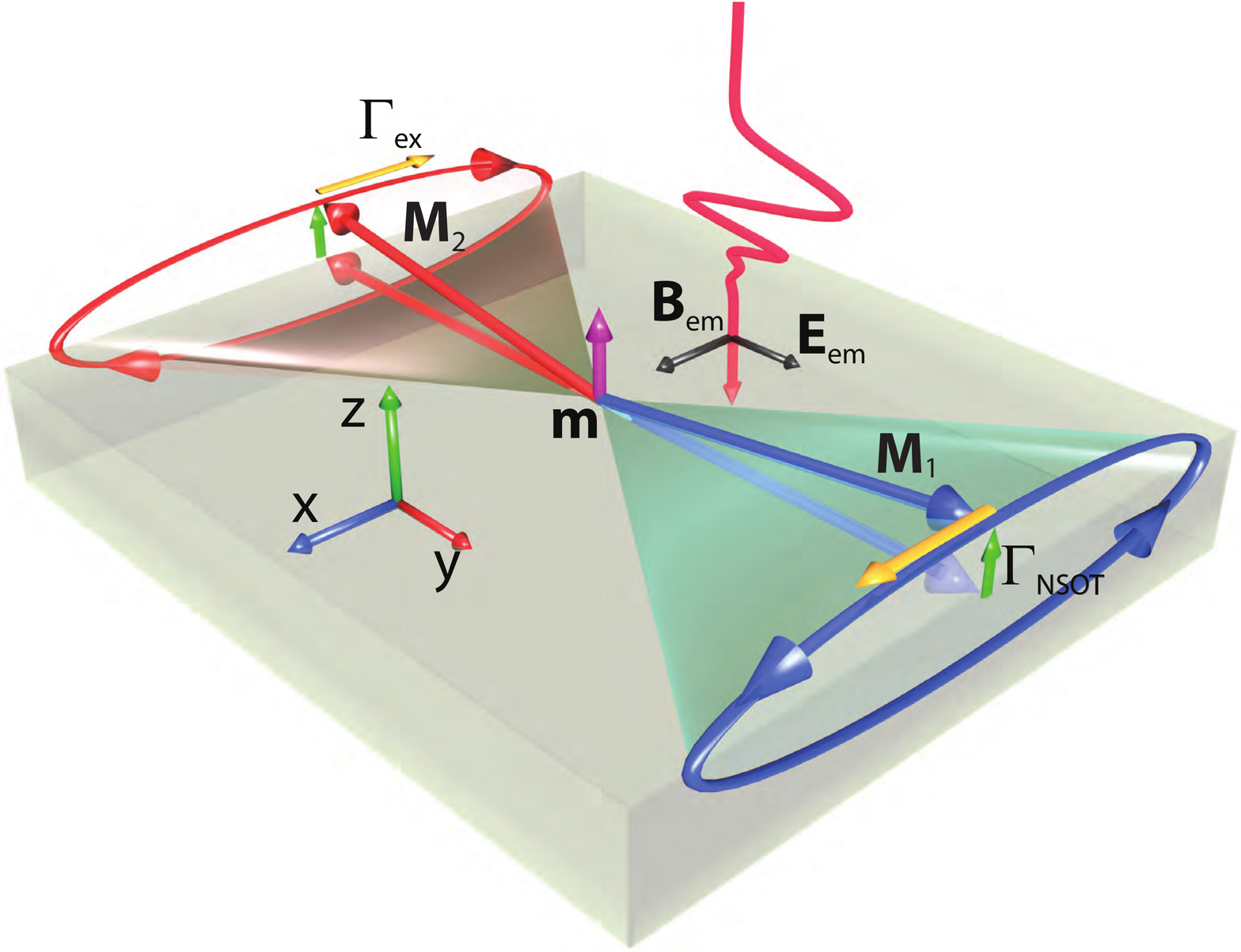}\caption{Schematic of the in-plane AFMR
mode driven by the in-plane polarized electromagnetic field pulse:
$\mathbf{M}_{1}$ and $\mathbf{M}_{2}$ are the sub-lattice magnetizations,
$\mathbf{m}$ is the resulting (oscillating) net magnetization. Due to the lack
of the local inversion symmetry in Mn$_{2}$Au the non-spin polarized currents
produce a staggered NSOT field \cite{Zelezny}. The resulting torques,
$\Gamma_{\mathrm{NSOT}}$, cant the two sub-lattice magnetizations
out-of-plane, where the torque resulting from a strong exchange field,
$\Gamma_{\mathrm{ex}}$, drives the in-plane precession.}%
\end{figure}

The experimentally determined $\nu_{\text{{\tiny AFMR}}}$ should be compared
to the expected eigenmode frequencies of Mn$_{2}$Au. Since Mn$_{2}$Au is an
easy plane AFM, with a strong out-of-plane and very weak in-plane magnetic
anisotropy, two nearly linearly polarized AFMR modes exist, similar to NiO
\cite{Tinkham}. The first mode, sketched in Figure 4 is an in-plane
($\parallel$) mode, where the two sub-lattice magnetization vectors,
$\mathbf{M}_{1}$ and $\mathbf{M}_{2}$, with $\left\vert \mathbf{M}%
_{1}\right\vert =\left\vert \mathbf{M}_{2}\right\vert =M_{s}$, precess in
opposite directions. The bases of the resulting cones are narrow ellipses
whose long axes lie in the easy plane. While in equilibrium $\mathbf{M}_{1}$
and $\mathbf{M}_{2}$ are anti-parallel and fully compensate each other, the
dynamics result in a small oscillating net magnetization $\mathbf{m}%
=\mathbf{M}_{1}+\mathbf{M}_{2}\neq0$ pointing in the direction of the hard
axis (Figure 4). The second eigenmode is the out-of-plane ($\perp$) mode with
the long axes of ellipses pointing along the hard-axis ($\widehat{z}$) and the
oscillating net magnetization lying in the easy-plane.

Following Kittel's approach \cite{Kittel}, we estimate the two
eigenfrequencies of Mn$_{2}$Au as $\omega_{\text{{\tiny {AFMR}}}}%
^{\parallel,\perp}={\gamma}\sqrt{H_{\mathrm{ex}}H_{\mathrm{an}}^{\parallel
,\perp}}$, where $H_{\mathrm{ex}}$ is the exchange field, $H_{\mathrm{an}%
}^{\parallel}$ and $H_{\mathrm{an}}^{\perp}$ are the in-plane and the
out-of-plane anisotropy field, respectively, and $\gamma=1.76\cdot10^{11}$
s$^{-1}$ T$^{-1}$ is the gyro-magnetic ratio. Using $H_{\mathrm{ex}}%
\approx1300$ T \cite{Barthem}, $H_{\mathrm{an}}^{\parallel}\approx0.3$ T, and
$H_{\mathrm{an}}^{\perp}\approx10$ T \cite{Shick} we obtain $\nu
_{\text{{\tiny AFMR}}}^{\parallel}\approx0.6$ THz and $\nu_{\text{{\tiny AFMR}%
}}^{\perp}\approx5$ THz, where $\nu_{\text{{\tiny {AFMR}}}}^{\parallel,\perp
}\equiv\omega_{\text{{\tiny {AFMR}}}}^{\parallel,\perp}/2\pi$. These estimates
suggest that we observe the in-plane AFMR (Fig. 4).

As noted above, the mode's absorption strength is much higher than in
insulating AFMs like NiO \cite{Tinkham,Kampfrath}, MnO \cite{Tinkham} and
$\alpha$-RuCl$_{3}$ \cite{THzAFMR}, where sample thicknesses of several 100
$\mu$m are required for THz absorption measurements
\cite{Tinkham,Kampfrath,THzAFMR,TGG,Hematite}. For quantitative assessment, we
compare the THz absorption coefficients at the center-frequencies of AFMRs,
$\alpha_{\text{{\tiny AFMR}}}$, for several AFMs. In Mn$_{2}$Au the
coefficient is $\alpha_{\text{{\tiny AFMR}}}\approx$ 880 mm$^{-1}$ at 50 K and
300 mm$^{-1}$ at 300 K, in NiO $\alpha_{\text{{\tiny AFMR}}}\approx0.18$
mm$^{-1}$ at 300 K \cite{Gonokami} and in $\alpha$-RuCl$_{3}$ $\alpha
_{\text{{\tiny AFMR}}}\approx0.2$ mm$^{-1}$ at 4 K \cite{THzAFMR}.

In insulating AFMs, like NiO, the Zeeman torque exerted by the magnetic field
component of the THz pulse is commonly ascribed to drive\ the AFMR mode
\cite{Kittel,Tinkham,Kampfrath}. However, the in-plane AFMR in Mn$_{2}$Au does
not couple to the magnetic field component of the THz pulse (in-plane
polarized). This fact, together with the three orders of magnitude larger
absorption compared to insulating AFMs, indicates that an alternative driving
mechanism of the AFMR in Mn$_{2}$Au must be present. Recent reports on current
induced switching of the N\'{e}el vector in both CuMnAs
\cite{Wadley,WadleyNew} and Mn$_{2}$Au \cite{Bodnar,Meinert}, where in the
former even switching with intense THz pulses was demonstrated \cite{Olejnik},
demonstrate the effectiveness of NSOT.

Next we analyze if this same scenario involving NSOT is able to quantitatively
account for both, the polarization and strength of the observed AFMR. Due to
the strong exchange coupling between the two magnetic sub-lattices, given by
$H_{\mathrm{ex}}$, the oscillating net magnetization $\mathbf{m}$ of the AFMR
is very small and the dynamics can be fully described by the N\'{e}el vector,
$\mathbf{n}=\mathbf{M}_{1}-\mathbf{M}_{2}$. Interaction of a system with
broken local inversion symmetry, like CuMnAs and Mn$_{2}$Au, with an
electromagnetic wave results in two external field-like torques: i) a torque
created by a time-dependent magnetic field component, $\propto\mathbf{n}%
\times\dot{\mathbf{B}}_{\mathrm{em}}\times\mathbf{n}$ \cite{Satoh}, and ii) a
N\'{e}el spin orbit torque \cite{Zelezny,Helen} $\propto\mathbf{n}%
\times\mathbf{\hat{z}}\times\mathbf{E}_{\mathrm{em}}$, which is driven by
current $\mathbf{j}=\sigma\mathbf{E}_{\mathrm{em}}$, induced by an electric
field component of an electro-magnetic wave, $\mathbf{E}_{\mathrm{em}}$, and
$\sigma$ is the optical conductivity. The resulting equation of motion for the
N\'{e}el vector is given by \cite{Helen,Helen2}:
\begin{align}
&  \mathbf{n}\times(\ddot{\mathbf{n}}+2\alpha_{G}\gamma H_{\mathrm{ex}}%
\dot{\mathbf{n}}-2\gamma^{2}H_{\mathrm{ex}}M_{s}\mathbf{H}_{\mathbf{n}%
})\nonumber\\
&  =\gamma\mathbf{n}\times(\dot{\mathbf{B}}_{\mathrm{em}}\times\mathbf{n}%
+2\lambda_{\mathrm{NSOT}}\sigma H_{\mathrm{ex}}M_{s}\mathbf{E}_{\mathrm{em}%
}\times\hat{z}).
\end{align}
Here $\alpha_{G}$ is the Gilbert damping constant, $\mathbf{H}_{\mathbf{n}%
}=-\partial w_{\mathrm{an}}/\partial\mathbf{n}$ is the internal effective
field, determined by the magnetic anisotropy energy landscape, $w_{\mathrm{an}%
}$, and $\lambda_{\mathrm{NSOT}}$ is a constant proportional to the NSOT
strength \cite{Helen2}. With the electromagnetic wave polarized within the
$x-y$ plane, where $\mathbf{E}_{\mathrm{em}}\Vert\mathbf{n}^{(0)}\Vert\hat{x}$
and $\mathbf{B}_{\mathrm{em}}\Vert\hat{y}$ (see Figure 4), the resulting
equations for small deviations of the N\'{e}el vector from its equilibrium
state, $\delta n_{y,z}$ (in dimensionless form), are%

\begin{align}
\delta\ddot{n}_{y}  &  +2\alpha_{\mathrm{G}}\gamma H_{\mathrm{ex}}\delta
\dot{n}_{y}+(\omega_{\text{{\tiny {AFMR}}}}^{\parallel})^{2}\delta
n_{y}\nonumber\\
&  =-\gamma\lambda_{\mathrm{NSOT}}\sigma H_{\mathrm{ex}}E_{\mathrm{em}},\\
\delta\ddot{n}_{z}  &  +2\alpha_{\mathrm{G}}\gamma H_{\mathrm{ex}}\delta
\dot{n}_{z}+(\omega_{\text{{\tiny {AFMR}}}}^{\perp})^{2}\delta n_{z}%
=-\gamma\dot{B}_{\mathrm{em}}.\nonumber
\end{align}

The above equations show that the NSOT drives the in-plane mode, while the
torque created by the magnetic field component of the THz pulse couples to the
out-of-plane mode. Moreover, taking the theoretically estimated $\lambda
_{\mathrm{NSOT}}=5-50$ s$^{-1}$A$^{-1}$cm$^{2}$ \cite{Zelezny,Zelezny2017},
$\omega_{\text{{\tiny {AFMR}}}}^{\parallel}\approx6.3\cdot10^{12}$ s$^{-1}%
\ $and $\sigma\left(  1\text{ THz, 4 K}\right)  \approx3.4\cdot10^{4}$%
~$\Omega^{-1}$cm$^{-1}$, and assuming $E_{\mathrm{em}}/B_{\mathrm{em}}=c$, a
comparison of the strengths of the two driving fields gives $\left(
2\lambda_{\mathrm{NSOT}}\sigma E_{\mathrm{em}}\right)  /B_{\mathrm{em}}%
\omega_{\text{{\tiny {AFMR}}}}^{\parallel}/\approx10^{2}-10^{3}$, accounting
for the anomalously large absorption strength in Mn$_{2}$Au compared to
insulating AFMs.

Based on the measured linewidth $\gamma_{\text{{\tiny AFMR}}}$ and
${\nu_{\text{{\tiny {AFMR}}}}^{\parallel}}$ and Eq.(2) we can also estimate
the value of the Gilbert damping. From ${\gamma_{\text{{\tiny AFMR}}}}%
/{\nu_{\text{{\tiny {AFMR}}}}^{\parallel}}=2\alpha_{G}\sqrt{H_{\mathrm{ex}%
}/H_{\mathrm{an}}^{\parallel}}$, where $\gamma_{\text{{\tiny AFMR}}}=0.3$~THz
and $\nu_{\text{{\tiny {AFMR}}}}^{\parallel}=1$~THz, we obtain a low value of
$\alpha_{G}\approx2.5\cdot10^{-3}$, which is typical for metallic
antiferomagnets \cite{Helen2}.

Finally, from the experimentally determined frequencies we obtain the
spin-flop field, $H_{\mathrm{s-f}}\equiv\omega_{\text{{\tiny {AFMR}}}%
}^{\parallel}/{\gamma}=\sqrt{H_{\mathrm{ex}}H_{\mathrm{an}}^{\parallel}}$, to
be 35 T at 4 K and 30 T at 300 K, consistent with recent measurements on
identical films exposed to pulsed magnetic fields \cite{Alexey}.

In summary, time-domain THz spectroscopy of Mn$_{2}$Au thin films reveals the
presence of a strong mode near 1 THz. The comparison of the mode's frequency
to the estimated AFMR frequencies of Mn$_{2}$Au shows that the mode is likely
an in-plane AFMR. Compared to previous reports on insulating AFMs, the mode
has an anomalously high absorption strength. Since the (in-plane) magnetic
field component of the THz pulse only weakly couples to the in-plane AFMR, we
suggest the in-plane AFMR in Mn$_{2}$Au is driven by the AC current producing
a N\'{e}el spin-orbit torque. The high frequency of the mode, its driving
mechanism, and the recently demonstrated DC-current switching of the N\'{e}el
vector \cite{Bodnar} make Mn$_{2}$Au a prime candidate for AFM ultrafast
memory applications.

\begin{acknowledgments}
This work was supported by the DFG in the framework of the Collaborative
Research Centre SFB TRR 173 \textquotedblleft Spin + X\textquotedblright. O.G
and J.S. acknowledge support from the Alexander von Humboldt Foundation, the
Grant Agency of the Czech Republic grant no. 14-37427G, the DFG (project SHARP
397322108), and the EU FET Open RIA Grant no. 766566. We acknowledge
discussions with T. Kampfrath, R. V. Mikhaylovskiy, and Th. Rasing.
\end{acknowledgments}

\end{document}